\documentclass[useAMS,usenatbib]{mn2e}

\def\var{\hbox{V\,393\,Sco~}}

\usepackage{graphicx}
 \usepackage{times}
 \usepackage{lscape}

\title{The evolution stage and massive disc of the interacting binary V\,393 Scorpii}
\author[ Mennickent et al.]
  {R.E. Mennickent$^{1}$\thanks{E-mail: rmennick@astro-udec.cl},   
   G. Djura\v{s}evi\'c$^{2,3}$,
    Z. Ko{\l}aczkowski$^{1,4}$,  
    G. Michalska$^{1,4}$  \\
  $^1$Universidad de Concepci\'on, Departamento de Astronom\'{\i}a,
      Casilla 160-C, Concepci\'on, Chile\\
  $^{2}$ Astronomical Observatory, Volgina 7, 11060 Belgrade 38, Serbia   \\
 $^{3}$ Isaac Newton Institute of Chile, Yugoslavia Branch\\
    $^{4}$  Instytut Astronomiczny Uniwersytetu Wroclawskiego, Kopernika 11, 51-622 Wroclaw, Poland
 \\
  }
\date{}

\pagerange{\pageref{firstpage}--\pageref{lastpage}} \pubyear{2008}

\def\LaTeX{L\kern-.36em\raise.3ex\hbox{a}\kern-.15em
    T\kern-.1667em\lower.7ex\hbox{E}\kern-.125emX}

\begin{document}

\label{firstpage}

\maketitle

\begin{abstract} 
V\,393 Scorpii is a bright Galactic Double Periodic Variable showing a long photometric cycle of $\approx$ 253 days.  We present new $VIJK$ photometric time series for V\,393 Scorpii along with the analysis of ASAS $V$-band photometry. We disentangled all light curves into the orbital and long cycle components. The  ASAS $V$-band {\it orbital} light curve was modeled with two stellar components  plus a circumprimary optically thick  disc assuming a semidetached configuration. We present the results of this calculation, giving physical parameters for the stars and the disc, along with general system dimensions. Our results are in close agreement with those previously found by Mennickent et al. (2010) from IR spectroscopy and the modeling of the spectral energy distribution. The stability of the orbital light curve suggests that the stellar $+$ disc configuration remains stable during the long cycle. Therefore, the long cycle should be produced by an additional variable and not-eclipsed emitting structure. 
We discuss the evolutionary stage of the system finding the best match with one of the  evolutionary models of van Rensbergen et al. (2008).  According to these models, the system is found  to be after an episode of fast mass exchange that transferred 4 
M$_{\odot}$ from the donor to the gainer  in a period of 400.000 years. We argue that a significant fraction of this mass has not been accreted by the gainer but remains in an optically thick {\it massive} ($\sim$ 2 M$_{\odot}$) disc-like surrounding pseudo-photosphere  whose luminosity is not driven by viscosity but probably by reprocessed stellar radiation.
 Finally, we provide the result of our search for Galactic Double Periodic Variables and briefly discuss the outliers 
 $\beta$ Lyr and $RX$ Cas.

\end{abstract}

\begin{keywords}
stars: early-type, stars: evolution, stars: mass-loss, stars: emission-line,
stars: variables-others
\end{keywords}

\section{Introduction}
 
V\,393 Scorpii is one of the Galactic  Double Periodic Variables (DPVs),  a group of interacting binaries showing a long photometric cycle lasting roughly 33 times  the orbital period (Mennickent et al. 2003, Mennickent \& Ko{\l}aczkowski 2009, Michalska et al. 2009, Poleski et al. 2010). DPVs  have been interpreted as semi-detached interacting binaries   with ongoing cyclic episodes of mass loss into the interstellar medium  (Mennickent et al. 2008, Mennickent \& Ko{\l}aczkowski 2010). The 253-d long photometric cycle of V\,393\,Scorpii was discovered by Pilecki \&  Szczygiel (2007)  after inspection of the ASAS catalogue for eclipsing binaries with additional variability. The IUE-UV properties of V\,393 Scorpii were studied by Peters (2001) who found evidence for a hot temperature region produced by the tangential impact of the gas stream into the gainer photosphere. This region should be the origin of the superionized lines observed in the UV,  like N\,V, C\,IV and Si\,IV, that are likely produced by resonance scattering in a plasma of  temperature T $\sim$ 10$^{5}$ K and electron density $N_{e} \sim$  10$^{9}$ cm$^{-3}$ (Peters \& Polidan 1984). The star was also studied by means of  multi-epoch high-resolution  IR spectroscopy by Mennickent et al. (2010,  hereafter M10), who also studied broad-band photometry and IUE ultraviolet spectra. After summarizing the available literature of this object, these authors argued for a semidetached B3 + F0  binary with masses 8  M$_{\odot}$ and  2 M$_{\odot}$  for the gainer and donor  (hereafter also called  primary and secondary, respectively) and orbital separation of 35 $R_{\odot}$. Most remarkably, M10 found evidence  for large mass loss through the Lagrangian L3 point during epochs of long cycle minimum and claim that their observations  suggest that the mass loss producing the long cycle is probably concentrated in equatorial regions.  

In this paper we refine stellar and  system parameters of V\,393 Scorpii by fitting the light curve with a multicomponent model including a stationary circumprimary disc. 
A detailed report of the observations used in this paper is given in Section 2, our results are presented in Section 3, a detailed discussion of these results is given in Section 4 and our conclusions are presented in Section 5. 

\section{Observations}


We obtained  $V, I, J, K$ band  images with the robotic 60-cm REM  telescope operated at La Silla, Chile, during three seasons in 2008-2010. Bias and flat calibration images were regularly obtained,  and de-biasing and flat-fielding for all science images were done with standard photometric IRAF\footnote{IRAF is distributed by the National Optical Astronomy Observatories,
 which are operated by the Association of Universities for Research
 in Astronomy, Inc., under cooperative agreement with the National
 Science Foundation.} tasks.  Instrumental magnitudes were calculated using aperture photometry and also by applying the differential image analysis   technique (DIA, e.g. Allard \& Lupton 1998). Differential magnitudes were calculated between the target and nearby stars used as comparison stars. The DIA technique provided the less noisy light curves and are shown in this paper. We also analyzed $V$-band magnitudes available in the public ASAS archive\footnote{www.astrouw.edu.pl/asas/}. Typical  one-sigma error for ASAS and REM photometry is 0.03 and 0.05 mag,  respectively.  This is the kind of statistical error provided in the rest of the paper for some modeled or fitting quantities. A summary of all photometric observations analyzed in this paper is given in Table 1.

\begin{table}
\centering
 \caption{Summary of photometric data analyzed in this paper. 
 ASAS data can be found at  http://www.astrouw.edu.pl/asas/.
 }
 \begin{tabular}{@{}ccccc@{}}
 \hline
Observatory &Telescope &Filters &N  &HJD-range (-2450000)  \\
\hline   
La Silla &REM 60 cm& $V$ &359&4498.8391--5372.6688\\
La Silla &REM 60 cm& $I$  &340&4498.8402--5391.7978\\
La Silla &REM 60 cm& $J$ &210&4498.8393--5362.8939\\
La Silla &REM 60 cm& $K$ &169&4502.8541--5369.8908\\
Las Campanas&ASAS-array &V&903  &1950.8740--4973.8121\\
\hline
\end{tabular}
\end{table}


\begin{figure}
\scalebox{1}[1]{\includegraphics[angle=0,width=8.5cm]{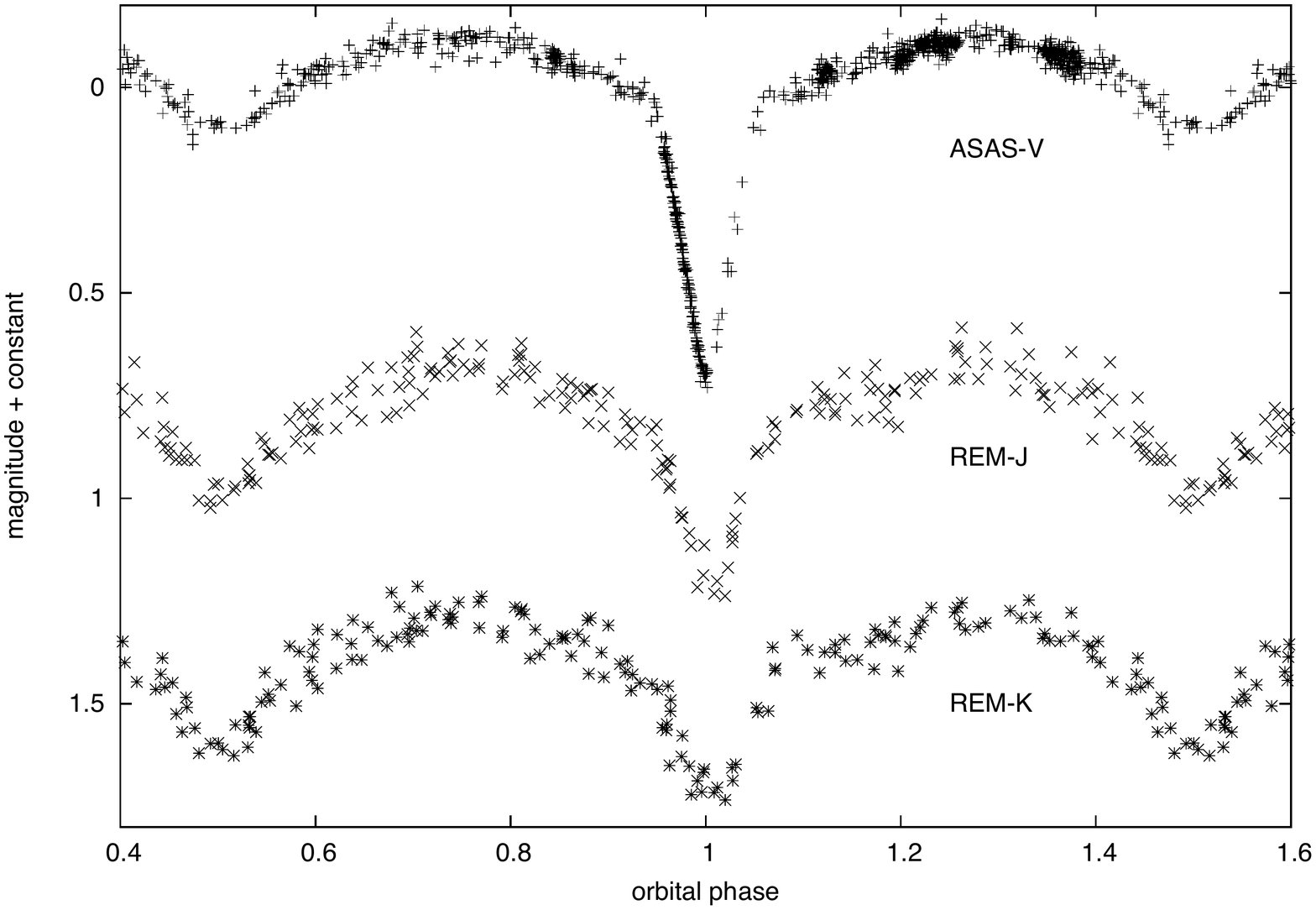}}
\scalebox{1}[1]{\includegraphics[angle=0,width=8.5cm]{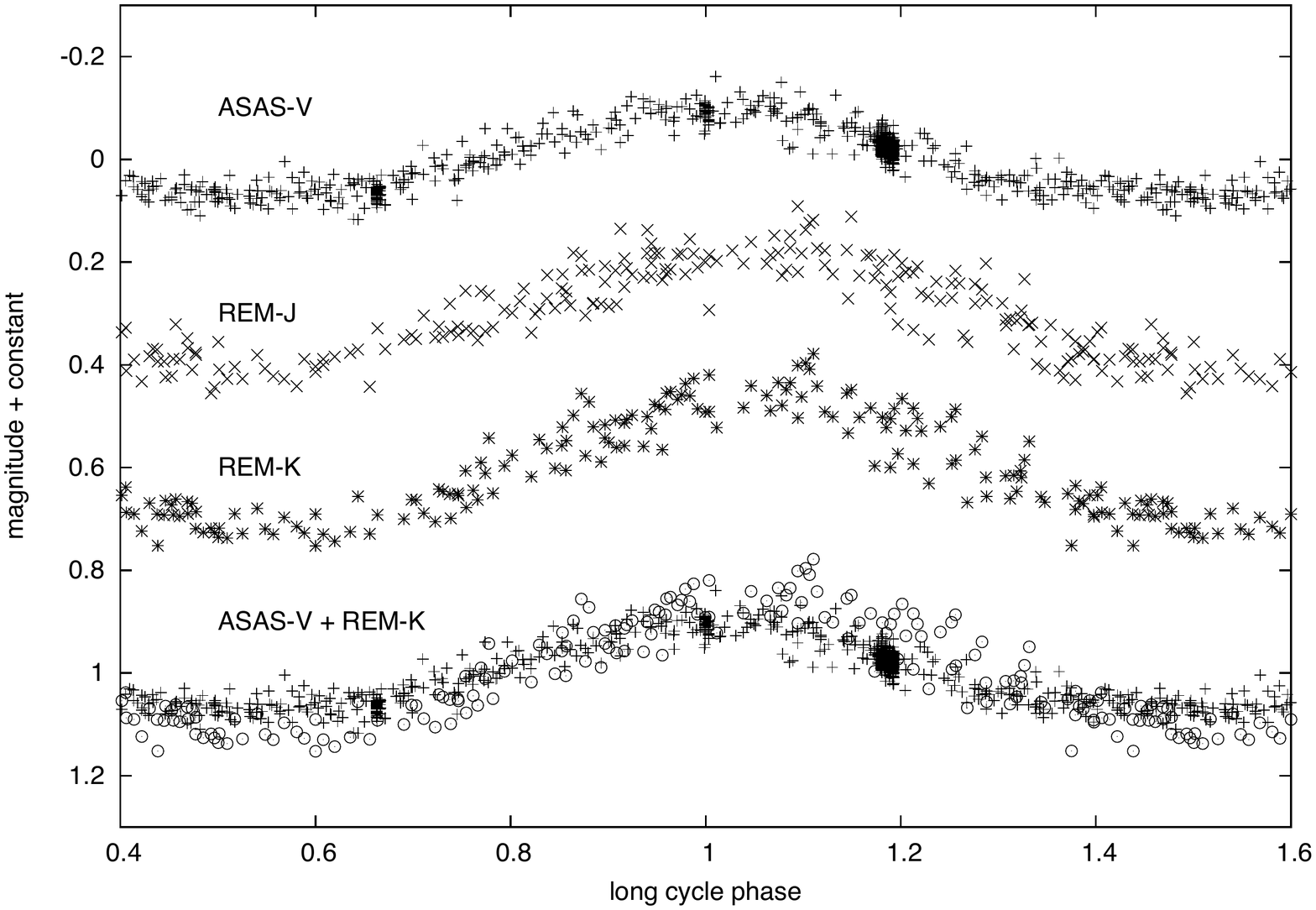}}
\caption{Orbital light curve (up) and long-cycle light curve (down) of V393 Scorpii at three bandpasses. Ephemeris are given in the text. In the lower light curve circles represent REM-$K$ and pluses ASAS-$V$ data.}
  \label{x}
\end{figure}

\section{Results}

\subsection{Light curve disentangling}


In this paper we use the ephemeris for the main orbital photometric minimum provided by Kreiner (2004), namely  $T_{o} = 2\,452\,507.7800 + 7.7125772 E$.  We separated the light curve of V393 Sco  into a long and a short period component. For that we used an algorithm especially designed to disentangle multiperiodic light curves through the analysis of their Fourier component amplitudes. 
The method  works as follows. A main frequency $f_{1}$ (usually the orbital one) is found with a period searching algorithm applied to the light curve. A least square fit is applied to the  light curve considering a fitting function consisting of a sum of sine functions of variable amplitude and phases representing the main frequency and
their additional significant harmonics. After this first fit, the residuals are inspected for a new periodicity $f_{2}$ (the long periodicity in DPVs). This new periodicity and their harmonics are included in the new fitting procedure. Finally, we obtain the
best representation for the light curve based on a sum of Fourier components of frequency $f_{1}$, $f_{2}$ and their significant harmonics. Data residuals with respect to the second and first theoretical light curves  form the orbital and long cycle photometric series, respectively.

This deconvolution method was applied to the light curves  of  V\,393 Sco at the observed bandpasses yielding  a unique solution consisting of  the orbital light curve plus  an additional smooth variability with period 253 days.  A phase dispersion minimization analysis (Stellingwerf 1978) of combined ASAS-$V$ and REM-$V$ data yielded for  the long-cycle the following ephemeris for the light curve maximum:\\

$HJD_{max, long} = 2452522 (\pm 1) + 253 (\pm 4) * E   \hfill(1)$\\

Inspection of the orbital and long light curves at different bandpasses reveals that (Fig.\,1): (i) the  rounded shape of the orbital light curves  at quadratures are consistent with a close binary showing proximity effects,  (ii) the system shows smooth long-cycle variability with larger amplitude in redder bandpasses. The amplitudes of this long variability in the $V, I, J$ and $K$ bands, derived from sinus fits to the REM data are 0.177 $\pm$ 0.003, 0.240 $\pm$ 0.004, 0.231 $\pm$ 0.003 and 0.263 $\pm$ 0.003 mag., respectively,  and (iii)   $V$ and $K$ light curves  look different, the maximum of the  latter seems to be delayed by $\Delta \Phi_{l} \approx$ 0.05  with respect to the former. 

The orbital light curve was examined with the program Period04\footnote{http://www.univie.ac.at/tops/period04/}. We calculated the error in the main frequency of  the fit to the orbital light curve.
The error consistently given by Monte Carlo simulations and the method of least squares is 4 $\times$ 10$^{-7}$ Hz.
This means that the period could drift at most by 4.76  $\times$ 10$^{-5}$ days in 3023 days (the ASAS-$V$ dataset time baseline). This implies that a constant orbital period change if present, should be less than 0.5 seconds per year.


\subsection{Model for a circumprimary optically thick disc}

Here we give a brief description of the  disc model that we apply to V\,393\,Sco. 

The basic elements of the binary system model with a plane-parallel  disc and the corresponding light-curve synthesis procedure are described in detail by Djura\u{s}evi\'{c} (1992, 1996).

We assume that the disc in V\,393 Sco is optically and geometrically thick. The disc edge is approximated by a cylindrical surface. In the current version of the model (Djura\u{s}evi\'{c}  et al. 2010), the thickness of the disc can change linearly with radial distance, allowing the disc to take a conical shape (convex, concave or plane-parallel). The geometrical properties of the disc are determined by its radius ($R_{d}$), its thickness at the edge ($d_{e}$) and the thickness at the center ($d_{c}$). 

The cylindrical edge of the disc is characterized by its temperature, $T_{d}$, and the conical surface of the disc by a radial temperature profile obtained by modifying the temperature distribution proposed by Zola (1991):\\

$T (r) = T_{d} + (T_{h} - T_{d}) [1 - (\frac{r - R_{h}}{ R_{d} - R_{h}})]^{a_{T}}$ \hfill(2) \\

We assume that the disc is in physical and thermal contact with the gainer, so the inner radius and temperature of the disc are equal to the temperature and radius of the star ($R_{h}$, $T_{h}$). The temperature of the disc at the edge ($T_{d}$) and the temperature exponent ($a_{T}$ ), as well as the radii of the star ($R_{h}$) and of the disc ($R_{d}$) are	free parameters, determined by solving the inverse problem.

The model of the system is refined by introducing active regions on the edge of the disc. 
The active regions have higher local temperatures so their inclusion results in a non-uniform distribution of radiation. The model includes  two such active regions: a hot spot (hs) and a bright spot (bs). These regions are characterized by their temperatures $T_{hs,bs}$ angular dimensions $\theta_{hs, bs}$ and longitudes $\lambda_{hs, bs}$.  These parameters are also determined by solving the inverse problem.

Due to the infall of an intensive gas-stream, the disc surface in the region of the hot spot becomes deformed as the material accumulates at the point of impact, producing a local deviation of radiation from the uniform azimuthal distribution.  
In the model, this deviation is described by the angle $\theta_{rad}$  between the line perpendicular to the local disc edge surface and the direction of the hot spot maximum radiation  in
the orbital plane.

The second spot in the model, i.e. the bright spot,  approximates the spiral structure of an accretion disc, predicted by hydrodynamical calculations (Heemskerk 1994). The tidal force exerted by the donor star causes a spiral shock, producing one or two extended spiral arms in the outer part of the disc. The bright spot can also be interpreted as a region where the disc significantly deviates from the circular shape.

\subsection{Results of the $V$-band light curve fitting}

The light-curve fitting  was performed using the inverse-problem solving method (Djura\u{s}evi\'{c}  1992) based on the simplex algorithm, and the model of a binary system with a disc described in the previous section. To obtain reliable estimates of the system parameters, a good practice is to restrict the number of free parameters by fixing some of them to values obtained from independent sources.
Thus we fixed the spectroscopic mass ratio to $q$ = 0.25  and the donor temperature  to $T_{2}$ = 7900 $K$, based on our spectroscopic study (Mennickent et al. in preparation).
In addition, we set the gravity darkening coefficient and the albedo of the gainer and the donor to $\beta_{h,c}$  = 0.25 and $A_{h,c}$ = 1.0 in accordance with von Zeipel's law for radiative shells and complete re-radiation  (Von Zeipel 1924). The limb-darkening for the components was calculated in the way described by Djura\u{s}evi\'{c}  et al.  (2010).

We treated the rotation of the donor as synchronous ($f_{c}$ = 1.0) since it is assumed that it has filled its Roche lobe (i.e. the filling factor of the donor was set to $F_{c}$ = 1.0). In the case of the gainer, however, the accreted material from the disc is expected to transfer enough angular momentum to increase the rate of the gainer up to the critical velocity as soon as even a small fraction of the mass has been transferred (Packet 1981, de Mink, Pols \& Glebbeek 2007). This means that the gainer fills its corresponding non-synchronous Roche lobe for  the star rotating in  the critical regime, and its dimensions and the amount of rotational distortion are uniquely determined by the factor of non-synchronous rotation, which is the ratio between the actual and the Keplerian angular velocity. For V\,393 Scorpii we assumed critical rotation for the gainer, i.e. $f_{h} = 14.6$ (Model A) but also calculated a model using synchronous rotation (Model B) to estimate the effect of the gainer velocity in the system parameters.  

The results of the light-curve analysis based on the described model of \var given in Table 2 basically confirm our solution for the stellar parameters given in M10,
based on IR spectroscopy and the modeling of the spectral energy distribution.  Furthermore, now we obtain a more realistic $R_{h}$ value. Our results obtained considering 
synchronous rotation for the gainer  do not differ much from the critical rotational case. The fit and stellar and disc dimensions are illustrated in Fig.\,2. We note that residuals show no dependence on orbital or long-cycle phases.

 We find that the best fit model of \var contains an optically and geometrically thick  disc around the hotter, more massive gainer star. With a radius of $R_{d} \approx 9.7 R_{\odot}$, the disc is more than twice as large as the central star ($R_{h} \approx 4.4 R_{\odot}$). The disc has a moderately convex shape, with central thickness of $d_{c} \approx 2.1R_{\odot}$ and the thickness at the edge of $d_{e} \approx 1.3 R_{\odot}$. The temperature of the disc increases from $T_{d}$ = 8600 $K$ at its edge, to $T_{h}$ = 16600 $K$ at the inner radius, where it is in thermal and physical contact with the gainer. The temperature profile exponent $a_{T}$ in Eq.\,1 is 4.5, this means that the effective temperature of the disc is significantly higher than the temperature at its edge.

We were able to model the asymmetry of the light curve very precisely by incorporating two regions of enhanced radiation on the  disc: the hot spot (hs), and the bright spot (bs). The hot spot is situated at longitude $\lambda_{hs} \approx$  320 degree, roughly between the components of the system, at the place where the gas stream falls onto the disc. The longitude $\lambda$ is measured clockwise (as viewed from the direction of the +Z-axis, which is orthogonal to the orbital plane) with respect to the line connecting the star centers (+X-axis), in the range 0-360 degrees. The temperature of the hot spot is approximately 20\% higher than the disc edge temperature, i.e. $T_{hs} \approx 10300$ $K$. The hot spot can be interpreted as a rough approximation of the Óhot lineÓ which forms at the edge of the gas stream between the components (Bisikalo et al. 2003).  Although including the hot spot region into the model significantly improves the fit, it cannot explain the light- curve asymmetry completely. By introducing one additional bright spot, larger than the hot spot and located on the disc edge at  $\lambda_{bs} \approx$  160 degree, the fit becomes much better.


\begin{figure}

\includegraphics[]{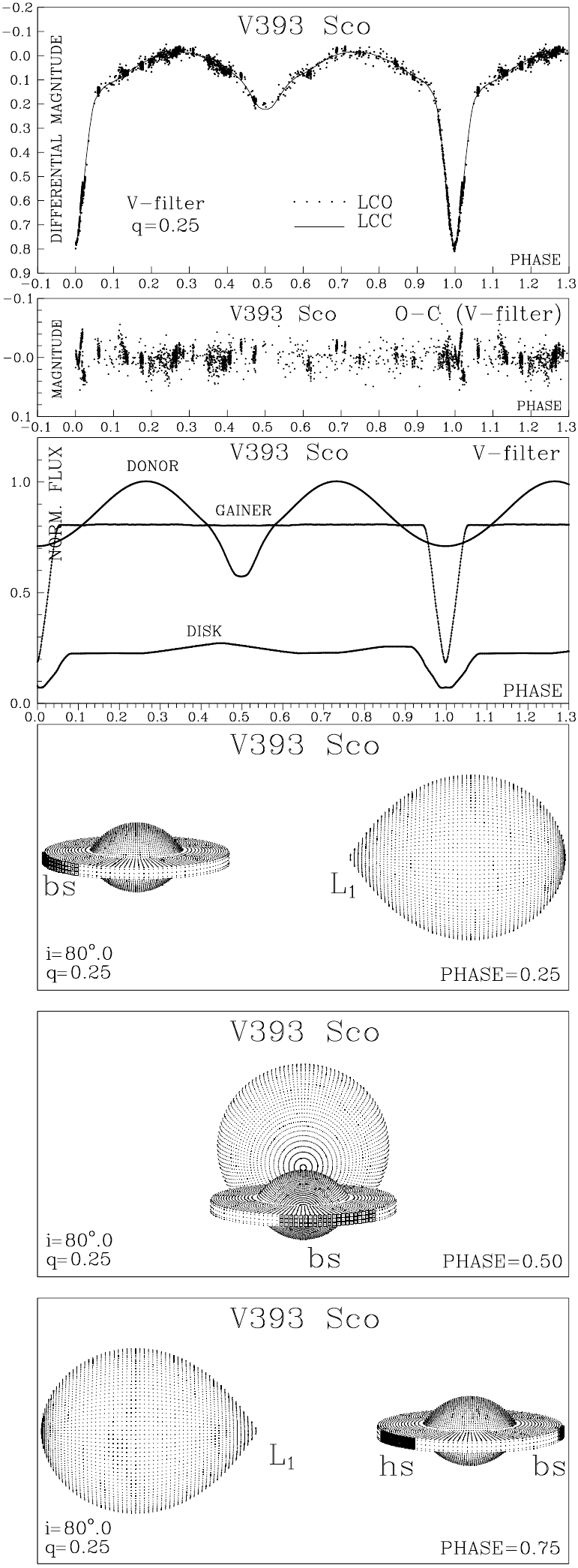}
\caption{Observed (LCO) and synthetic (LCC) light-curves of
V393 Sco obtained by analyzing photometric observations; final
O-C residuals between the observed and optimum synthetic light curves;
fluxes of donor, gainer and of the  disc, normalized
to the donor flux at phase 0.25; the views
of the optimal model at orbital phases 0.25, 0.50 and 0.75,
obtained with parameters estimated by the light curve analysis.}
\label{fV393Sco}
\end{figure}

\begin{table*}

\caption{Results of the analysis of {V393 Sco} V-filter light-curve
obtained by solving the inverse problem for the Roche model with
a disc around the more-massive (hotter) component in
critical rotation regime (Model A) and synchronous rotational regime (Model B).}

 \label{TabV393Sco}
      \[
        \begin{array}{llllllll}
            \hline
            \noalign{\smallskip}

{\rm Quantity} &{\rm Model~A} & {\rm Quantity} &{\rm Model~A} &{\rm Quantity} &{\rm Model~B}& {\rm Quantity} &{\rm Model~B}\\
            \noalign{\smallskip}
            \hline
            \noalign{\smallskip}
   n                                       & 2268                   & \cal M_{\rm_h} {[\cal M_{\odot}]} & 7.8  \pm 0.2  &             n                      & 2268                            & \cal M_{\rm_h} {[\cal M_{\odot}]} &7.8 \pm 0.5   \\
{\rm \Sigma(O-C)^2}                & 0.5638              & \cal M_{\rm_c} {[\cal M_{\odot}]} & 2.0  \pm 0.2  &{\rm \Sigma(O-C)^2}     & 0.5621     & \cal M_{\rm_c} {[\cal M_{\odot}]} & 2.0 \pm 0.2 \\
{\rm \sigma_{rms}}                 & 0.0157                & \cal R_{\rm_h} {\rm [R_{\odot}]}  & 4.4  \pm 0.2  &{\rm \sigma_{rms}}          &  0.0157   &\cal R_{\rm_h} {\rm [R_{\odot}]}  & 4.1 \pm 0.2\\
   i {\rm [^{\circ}]}                   & 80.0  \pm 0.2     & \cal R_{\rm_c} {\rm [R_{\odot}]}  & 9.4  \pm 0.3  &  i {\rm [^{\circ}]}          &   79.9 \pm 0.2    &\cal R_{\rm_c} {\rm [R_{\odot}]}  &  9.4 \pm 0.3 \\
{\rm F_d}                               & 0.55  \pm 0.04   & {\rm log} \ g_{\rm_h}                  & 4.0  \pm 0.1  &{\rm F_d}                        &   0.51 \pm 0.04 &{\rm log} \ g_{\rm_h}                  &  4.1 \pm 0.1\\
{\rm T_d} [{\rm K}]                & 8600  \pm 600  & {\rm log} \ g_{\rm_c}                   & 2.8  \pm 0.1  &{\rm T_d} [{\rm K}]         &   8300 \pm 600 &{\rm log} \ g_{\rm_c}                   & 2.8 \pm 0.1\\
{\rm d_e} [a_{\rm orb}]            & 0.04 \pm 0.01  & M^{\rm h}_{\rm bol}                    &-3.0  \pm 0.2  &{\rm d_e} [a_{\rm orb}]    &   0.04 \pm 0.01  &M^{\rm h}_{\rm bol}                    & -2.9 \pm 0.3        \\
{\rm d_c} [a_{\rm orb}]            & 0.06 \pm 0.01  & M^{\rm c}_{\rm bol}                     &-1.4  \pm 0.1  &{\rm d_c} [a_{\rm orb}]     &  0.08 \pm 0.01  &M^{\rm c}_{\rm bol}                     & -1.4 \pm 0.1 \\
{\rm a_T}                               & 4.5   \pm 0.5     & a_{\rm orb}  {\rm [R_{\odot}]}     & 35.1 \pm 0.3  &{\rm a_T}                       &    4.4 \pm 0.5  &a_{\rm orb}  {\rm [R_{\odot}]}     &  35.1 \pm 0.5    \\
{\rm f_h}                               & 14.6  \pm 0.8     & \cal{R}_{\rm d} {\rm [R_{\odot}]}  & 9.7  \pm 0.3  &{\rm f_h}                       &    1.00   &\cal{R}_{\rm d} {\rm [R_{\odot}]}  & 9.1 \pm 0.5\\
{\rm F_h}                               & 1.00                 & \rm{d_e}  {\rm [R_{\odot}]}          & 1.3  \pm 0.3  &{\rm F_h}                         &  0.25 \pm 0.01&  \rm{d_e}  {\rm [R_{\odot}]}          & 1.4 \pm 0.3       \\
{\rm T_h} [{\rm K}]                & 16600 \pm 500 & \rm{d_c}  {\rm [R_{\odot}]}         & 2.1  \pm 0.4  &{\rm T_h} [{\rm K}]     &       16800 \pm 500 &\rm{d_c}  {\rm [R_{\odot}]}         &  2.8 \pm 0.5          \\
{\rm A_{hs}=T_{hs}/T_d}                & 1.2    \pm 0.1  &         &     &       {\rm A_{hs}=T_{hs}/T_d}                &    1.3  \pm 0.1   &      &       \\
{\rm \theta_{hs}}{\rm [^{\circ}]}       & 19.0   \pm 2.0   &     &     &        {\rm \theta_{hs}}{\rm [^{\circ}]}       &    18.5   \pm   2.0 &  &       \\
{\rm \lambda_{hs}}{\rm [^{\circ}]}    & 324.0 \pm 6.0   &    &      &       {\rm \lambda_{hs}}{\rm [^{\circ}]}    &    325.0  \pm   6.0  & &           \\
{\rm \theta_{rad}}{\rm [^{\circ}]}     & -5.0   \pm 5.0   &     &     &        {\rm \theta_{rad}}{\rm [^{\circ}]}     &    -5.0   \pm  5.0  &   &             \\
{\rm A_{bs}=T_{bs}/T_d}              & 1.2      \pm 0.1  &         &     &         {\rm A_{bs}=T_{bs}/T_d}              &   1.3   \pm   0.1  &     &             \\
{\rm \theta_{bs}} {\rm [^{\circ}]}     & 33.0   \pm 6.0   &     &     &        {\rm \theta_{bs}} {\rm [^{\circ}]}     &   29.0    \pm   6.0  &  &                      \\
{\rm \lambda_{bs}}{\rm [^{\circ}]}   & 162.0  \pm 9.0   &    &     &       {\rm \lambda_{bs}}{\rm [^{\circ}]}   &    154.0 \pm    9.0   &   &                         \\
{\Omega_{\rm h}}                       & 9.90     \pm 0.03  &     &     &          {\Omega_{\rm h}}                       &    8.83  \pm   0.03     &      &                  \\
{\Omega_{\rm c}}                        & 2.35      \pm 0.02  &     &     &           {\Omega_{\rm c}}                        & 2.35    \pm   0.02      &      &                    \\
            \noalign{\smallskip}
            \hline
         \end{array}
      \]

FIXED PARAMETERS: $q={\cal M}_{\rm c}/{\cal M}_{\rm
h}=0.25$ - mass ratio of the components, ${\rm T_c=7900 K}$  -
temperature of the less-massive (cooler) donor, ${\rm F_c}=1.0$ -
filling factor for the critical Roche lobe of the donor,
$f{\rm _c}=1.00$ - non-synchronous rotation coefficients
of the donor, ${\rm \beta_{h,c}=0.25}$
- gravity-darkening coefficients of the components, ${\rm
A_{h,c}=1.0}$  - albedo coefficients of the components.

\smallskip \noindent Note: $n$ - number of observations, ${\rm
\Sigma (O-C)^2}$ - final sum of squares of residuals between
observed (LCO) and synthetic (LCC) light-curves, ${\rm
\sigma_{rms}}$ - root-mean-square of the residuals, $i$ - orbit
inclination (in arc degrees), ${\rm F_d=R_d/R_{yc}}$ - disc
dimension factor (the ratio of the disc radius to the critical Roche
lobe radius along y-axis), ${\rm T_d}$ - disc-edge temperature,
$\rm{d_e}$, $\rm{d_c}$,  - disc thicknesses (at the edge and at
the center of the disc, respectively) in the units of the distance
between the components, $a_{\rm T}$ - disc temperature
distribution coefficient, $f{\rm _h}$ - non-synchronous rotation coefficient
of the more massive gainer (ratio between the rotational angular velocity and the synchronous rotational angular velocity),
${\rm F_h}=R_h/R_{zc}$ - filling factor for the
critical Roche lobe of the hotter, more-massive gainer (ratio of the
stellar polar radius to the critical non-synchronous Roche lobe
radius along z-axis for a star in critical rotation regime, equal to 1 for critical rotation regime),
${\rm T_h}$ - temperature of the gainer, ${\rm
A_{hs,bs}=T_{hs,bs}/T_d}$ - hot and bright spots' temperature
coefficients, ${\rm \theta_{hs,bs}}$ and ${\rm \lambda_{hs,bs}}$ -
spots' angular dimensions and longitudes (in arc degrees), ${\rm
\theta_{rad}}$ - angle between the line perpendicular to the local
disc edge surface and the direction of the hot-spot maximum
radiation, ${\Omega_{\rm h,c}}$ - dimensionless surface potentials
of the hotter gainer and cooler donor, $\cal M_{\rm_{h,c}} {[\cal
M_{\odot}]}$, $\cal R_{\rm_{h,c}} {\rm [R_{\odot}]}$ - stellar
masses and mean radii of stars in solar units, ${\rm log} \
g_{\rm_{h,c}}$ - logarithm (base 10) of the system components
effective gravity, $M^{\rm {h,c}}_{\rm bol}$ - absolute stellar
bolometric magnitudes, $a_{\rm orb}$ ${\rm [R_{\odot}]}$,
$\cal{R}_{\rm d} {\rm [R_{\odot}]}$, $\rm{d_e} {\rm [R_{\odot}]}$,
$\rm{d_c} {\rm [R_{\odot}]}$ - orbital semi-major axis, disc
radius and disc thicknesses at its edge and center, respectively,
given in solar units.

\end{table*}

 \section{Discussion}
 
 \subsection{Disc formation}
 
 In a semidetached system with the donor transferring matter onto the gainer through an accretion stream, the stream  has a distance of closest approach $r_{min}$ from the center of the gainer, given approximately by:\\
 
 $\frac{r_{min}}{a} = 0.0488q^{-0.464} $\hfill(3)\\
 
 \noindent 
 (Lubow \& Shu 1975).  Using $q$ = 0.25 we obtain $r_{min}$ = 0.093$a$, comparable to the gainer radius (0.125 $\pm$ 0.046 $a$). The fact that our light curve model shows a relatively large disc around the gainer indicates that: (i) the stream in principle hits the gainer that rapidly reaches critical rotation (Packet 1981),  it cannot accrete additional mass and a disc is built around the star or (ii) the gainer radius is smaller  than $r_{min}$ and a disc is formed naturally. At present, uncertainties in $q$ and $R_{h}$ do not allow to select between these mutually exclusive alternatives.

\subsection{On the cause for the long-cycle} 
 
 It is notable that the model containing the circumprimary disc fits very well the orbital light curve {\it through the whole long cycle}. This fact suggests that the optically thick disc does not participate in the long cycle, at least that part of the disc responsible for the orbital photometric variability. We conclude that the long cycle is produced by an additional variable  and non-eclipsed emitting  structure, a fourth light in the system. In addition, the fact that this fourth light is redder at long maximum and bluer at minimum places strong constrains on the possible cause for the long cycle variability.
 
In principle, it is possible that the long minimum is due to obscuration of the system by ejected material through the equatorial plane, as suggested by M10's observations. This possibility is not supported by our observations, since it should produce a white (opaque material) or red (gas producing reddening in the line of sight and free-free emission outside the line of sight) minimum.  On the contrary, our observations indicate that the system when brighter is redder,  a fact  consistent with free-free emission in non-obscured ejected circumbinary material. If this material is not ejected in the orbital plane at long maximum, then it is ejected at higher latitudes. In this sense the inferred variable emitting structure is reminiscent of the jets found in $\beta$ Lyr (Harmanec et al. 1996, Harmanec 2002.)

This interpretation conflicts with the M10 suggestion that mass loss occurs in the equatorial plane. That position was based on the non-variability of UV spectral features through the long cycle and the evidence for mass flows through the L3 point.  However, only few UV spectra were available for analysis, and it is possible that the putative variability was hidden by additional orbital variability, or alternatively, the long cycle variability disappears at UV wavelengths. Much more well-sampled UV spectra are needed to illuminate this point. In addition, the inferred mass loss through L3 around secondary eclipse is still compatible with vertical outflows. Furthermore, UV lines {\it do} show evidence for such high latitude outflows (M10).

 \subsection{Mass transfer and mass loss}
 
In this section we calculate the mass transfer rate ($\dot{M}$)  for Model A using several approaches. The radial temperature structure of a steady state accretion disc is given by:\\

$T(r) = T_{0} [\frac{r}{R_{h}}]^{-3/4} [1-[\frac{R_{h}}{r}]^{1/2}]^{1/4} $\hfill(4)\\

\noindent
where\\

$T_{0} = [\frac{3GM_{h}\dot{M}}  
{8\pi \sigma R_{h}^{3}} ]^{1/4}$
\hfill(5)\\

\noindent
where $G$ is the gravitational constant and $\sigma$ the Stephan-Boltzman constant (e.g. Warner 1995). From the above expressions and using the parameters given in Table 2 and $T$ = 8600 $K$ (the temperature in the disc outer edge) we obtain $\dot{M}$ = 
3.74 $\times$ 10$^{-5}$ M$_{\odot}$ yr$^{-1}$. 

Let's  use the formula given by Smak (1989) for the geometrical thickness $H$ of an $\alpha$-disc at the density level of 10$^{-10}$ g cm$^{-3}$:\\

$\frac{H}{R_{d}} 
\approx 0.07
[\dot{M}_{18} (1-(\frac{R_{h}}{R_{d}})^{0.5})]
^{0.18} $\hfill(6)\\

\noindent
where $\dot{M}_{18}$ is the mass accretion rate in units of  10$^{18}$ g s$^{-1}$. Using the average between $d_{e}$ and $d_{c}$ as the value for $H$, we found 
 $\dot{M}$ = 
7.97 $\times$ 10$^{-6}$ M$_{\odot}$ yr$^{-1}$. 

Now we assume that at $\Phi_{o}$ = 0.25 the system magnitude is $V$= 7.609 (Table 4). From the light curve model the  disc flux contribution at this phase is 10\%.
Considering the reddening $E(B-V)$ = 0.13 and a distance of 523 pc (M10) we obtain $V^{disc}$ = 9.68 and $M^{disc}_{V}$ = 1.09. Using the bolometric correction $BC$ = -1.50 for $T$ = 17000 $K$, log $g$ = 3.0, and $Z$ = $Z_{\odot}$ (Lanz \& Hubeny 2007) we obtain $M^{disc}_{bol}$ = -0.41 corresponding to 115 $L_{\odot}$.
If this luminosity is powered by accretion, then:\\

$L^{disc} = \frac{GM_{h}\dot{M}}{2R_{h}} $\hfill(7)\\

\noindent
(e.g. Warner 1995). Using the above equation we derive $\dot{M}$ = 4.15 $\times$ 10$^{-6}$ M$_{\odot}$ yr$^{-1}$. 

If the DPV long cycles are due to recurrent episodes of systemic mass loss as suggested by Mennickent et al. (2008), and assuming no mass accumulation around the gainer in the long term, then $\Delta M$ = $\dot{M}$ $P_{long}$ are ejected from the system every long cycle. For the $\dot{M}$ values given above this means that $\Delta M$ between 5.4 $\times$ 10$^{-5}$ and 2.9 $\times$  10$^{-6}$ M$_{\odot}$ are ejected in every cycle. We find that in an accretion powered system, the amount of mass lost in every long cycle is not minor and should affect the system evolution considerably. 

 If the system loses mass due to a spherically symmetric wind that does not interact with the companion, then the orbital period changes by (e.g. Hilditch 2001):\\

$\frac{\dot{P_{o}}}{P_{o}} = \frac{-2\dot{M}_{h}}{M_{h}+M_{c}}$\hfill(8)\\ 

Using our  stellar masses and estimates for the mass loss rate given above we find changes in the orbital period of 1 to 10 seconds per year, easily detectable with the current astronomical instrumentation. The fact that this variability is not observed argues against an accretion disc whose luminosity is driven by viscosity but in favor of an extended photosphere radiating by reprocessed stellar radiation. This view is also supported by other evidence presented later in the next section.





\begin{table}
\centering
 \caption{The parameters of the van Rensbergen et 
al.  (2008) model that best fits the V\,393 Sco data. The 
hydrogen and helium core mass fractions  ($X_{c}$ and $Y_{c}$) are given for the cool and hot star.} 
 \begin{tabular}{@{}lclc@{}}
 \hline
{\rm quantity}  &{\rm value} &  {\rm quantity} &{\rm value} \\
 \hline
age & 7.00E7   yr  & period &7.713 d  \\
$M_{c}$  &2.11 M$_{\odot}$& $M_{h}$  &7.49 M$_{\odot}$ \\
$\dot{M_{c}}$ &-9.47E-9   M$_{\odot}$/yr &$\dot{M_{h}}$  & 9.47E-9    M$_{\odot}$/yr\\
log $T_{c}$ &3.92   $K$ &log $T_{h}$ &4.32   $K$\\
log $L_{c}$ &2.61 L$_{\odot}$ &log $L_{h}$ &3.39 L$_{\odot}$\\
$R_{c}$ &9.55  R$_{\odot}$ &$R_{h}$ &3.79   R$_{\odot}$  \\
$X_{cc}$ &0.05 &$X_{ch}$ &0.63  \\
$Y_{cc}$ & 0.93 &$Y_{ch}$ &0.35    \\   
\hline
\end{tabular}
\end{table}

\subsection{Evolutionary stage of V\,393 Scorpii}

The comparison of the stellar parameters with predictions of evolutionary models for single stars with solar metallicity indicates: (i) the donor is significantly inflated appearing as an over-luminous object for its mass and (ii) the gainer is under-luminous and under-heated for the given mass (Fig.\,3). 
The inflated cooler star is expected in a semidetached interacting binary
with an evolved donor whereas the under-luminous hotter star can reveal the presence of a massive circumprimary disc.  In fact, the loci occupied by the 7.8 M$_{\odot}$ gainer in the HR diagram corresponds to a slightly evolved 6 M$_{\odot}$ star. The figure $R_{h}$ = 4.4 R$_{\odot}$ is also compatible with a 6 M$_{\odot}$ star.
This could be  possible if a 2 M$_{\odot}$ optically-thick disc surrounds the gainer. In other words, in this view, our dynamical mass is overestimated by the presence of a massive disc\footnote{The DPV AU\,Mon suffers of the same problem.  Considering  $M_{1}$ = 7 M$_{\odot}$, $\log T_{h}$ = 4.20 and $\log L_{h}$ = 3.17  found by Djura{\v s}evi{\'c} et al. (2010) the gainer turns to be under-luminous.}

The average disc electron number density can be estimated from:\\

$n_{e} = \frac{M_{disc}}{\pi Hm_{e}(R_{d}^{2}-R_{h}^{2})}$\hfill(9)\\ 

\noindent
Using the values given in Table 2 and  $M_{disc}$ = 2 M$_{\odot}$ we derive $n_{e}$ = 3.2 $\times$ 10$^{25}$ cm$^{-3}$. This density is much larger than that found in a normal B-type photosphere and comparable with densities found in the stellar interior. 

An  alternative explanation involving a less massive disc is that the gainer light is  in some way obscured by the presence of the high latitude wind detected in the UV producing a smaller effective temperature and luminosity. Whereas the hypothesis of a massive disc explains the small $R_{h}$, the obscuration effect does not. 

The hypothesis of a self-gravitating massive disc (0.5 $M_{\odot}$) for the DPV-related system $\beta$ Lyr was considered by Wilson \& Terrell (1992) but critiqued by Hubeny et al. (1994) who pointed out that a massive-disc model requires unrealistically low viscosity, i.e. a large Reynolds number, and is likely to be dynamically unstable. All these authors assumed accretion-powered discs in their calculations, leaving the door open for a structure with a different mechanism for energy generation. 
Further theoretical work is necessary to explore  alternative physical scenarios  for massive discs around normal stars more deeply.



Now we consider predictions of binary evolution models including epochs of systemic mass loss. We inspected the 561 conservative and non-conservative evolutionary tracks  by van Rensbergen et 
al.  (2008) available at the Center de Donn\'ees Stellaires (CDS) looking for the best match for the system parameters found for V\,393 Sco.
Models with strong and weak tidal interaction were studied, although only the  latter ones should allow critical rotation of the gainer. 
A  multi-parametric fit was made with the  synthetic ($S_{i,j,k}$)  and observed ($O_{k}$) stellar parameters  mass, temperature, luminosity and radii, and the orbital period,  where $i$ (from 1 to 561) indicates the synthetic  model, $j$ the time $t_{j}$ and $k$  (from 1 to 9)  the  stellar or orbital parameter.
 Non-adjusted parameters were mass loss rate, Roche lobe radii, chemical composition, fraction of accreted mass lost by the system and age.
 For every synthetic model $i$ we calculated the quantity $\chi^{2}_{i,j}$ at every $t_{j}$ defined by:\\

$\chi^{2}_{i,j} \equiv (1/N) \Sigma_{k} w_{k}[(S_{i,j,k}-O_{k})/O_{k}]^{2} $\hfill(10) \\

\noindent
 where $N$ is the normalization factor and $w_{k}$ the statistical weight of the parameter $O_{k}$, calculated as:\\

$w_{k} = \sqrt{O_{k}/\epsilon(O_{k})} $\hfill(11) \\ 

\noindent
where $\epsilon(O_{k})$ is the error associated to the observable $O_{k}$.
The model with the minimum $\chi^{2}$ corresponds to the model with the best evolutionary history of V\,393\,Sco. The absolute minimum $\chi^{2}_{min}$  identifies the age of the system along with the theoretical stellar and orbital parameters. The high accuracy of the orbital period dominates the search for the best solution in a single evolutionary track, but the others parameters play a role when comparing tracks corresponding to different initial stellar masses. 

We find the absolute $\chi^{2}$ minimum in the weak  interaction model with initial masses of 6 and 3.6 M$_{\odot}$ and initial orbital period of 3 days. 
This model starts Roche Lobe overflow at about the same time that core hydrogen burning ends, so it is a borderline 
Case A/B (thanks to Nicki Mennekens for this insight). The averages of their parameters are shown in Table 3.

The corresponding evolutionary tracks for the primary and secondary stars are shown in Fig.\,4, along with the position for the
best model for V\,393 Sco. We observe a relatively good match for the donor star parameters but a mismatch with the gainer temperature and luminosity. 
The higher stellar temperature and luminosity indicated by the model could indicate that in practice the gainer has not accreted all the transferred mass but  part of the mass has been accumulated in a massive optically thick surrounding disc. Consequently, lower temperatures and luminosities are observed for the gainer, those corresponding to the actual stellar mass. This conclusion is consistent with our previous  discussion and was obtained in an independent way.

The best fit also indicates that  V\,393 Sco is found after a burst of mass transfer, the gainer having received 4 M$_{\odot}$ in a rapid burst lasting 400.000 years.  The donor is an inflated ($R_{c}$ = 9.6 R$_{\odot}$) and evolved 2 M$_{\odot}$ star with its core consisting of 93\% of helium. 
According to the best model the system has now an age of 7.00 $\times$ 10$^{7}$ yr and 
$\dot{M}$ =  9.47 $\times$ 10$^{-9}$  M$_{\odot}$ yr$^{-1}$ (Fig.\,5). 

We  notice that the mass transfer rate derived from model fitting is much smaller than those derived in Section 4.3. This could indicate that: (i)  $\dot{M}$ values  are still not well reproduced by the models or (ii) basic assumptions of Section 4.1 are invalid, like accretion powered disc luminosity  and the hypothesis of a steady state accretion disc. The fact that we obtain different values of $\dot{M}$ using the accretion-powered hypothesis  and that we do not observe orbital period changes argue for the second alternative. If viscous dissipation in an accretion disc is not the source of disc luminosity then other mechanisms should be invoked, like shocks produced  in impact regions or photoionization of circumstellar gas  and recombination of high energy photons produced in hot regions. If the disc is considered an extension of the gainer photosphere, it should fit its temperature at the inner region but  to show  cooler temperatures at the outer regions, as  indicated  by Eq.\,2. In this case we expect the spectral energy distribution be characterized by thermal radiation from an optically thick disc in thermal contact with the hotter star. In this view, it is the gainer that  maintains the hot disc, not viscous dissipation.

The low $\dot{M}$ model imposes certain problems for the accretion 
powered disc, but explains the absence of measured orbital period changes. On the other hand, it is not impossible to assume that low $\dot{M}$ values  maintain the long cycle. If we assume  $\dot{M}$ = 9 $\times$ 10$^{-9}$  M$_{\odot}$ yr$^{-1}$ then 6.2 $\times$ 10$^{-9}$  M$_{\odot}$
are ejected in steady state in a long cycle. The low $\dot{M}$  solution implies that V\,393\,Sco is in a relatively long-lasting evolutionary stage compared with the previous rapid mass transfer rate stage. This implies that a relatively large number of systems should be found showing the DPV phenomenon, as actually observed. 

If we  require the best fit model to match the high $\dot{M}$  value then we should look at the end of the rapid mass transfer burst (younger system with age 6.8 $\times$ 10$^{7}$ years) or at the beginning of the following mass transfer event (older system with age 7.55 $\times$ 10$^{7}$ years, Fig.\,5). In both cases we can get  $\dot{M}$  values of the order of 10$^{-6}$ M$_{\odot}$ yr$^{-1}$. However, these models have larger $\chi^{2}$ values than the preferred model by a factor of  10.
In both cases, our conclusion that a  massive structure surrounds the gainer remains unaltered.
 


\begin{figure}
\scalebox{1}[1]{\includegraphics[angle=0,width=8.5cm]{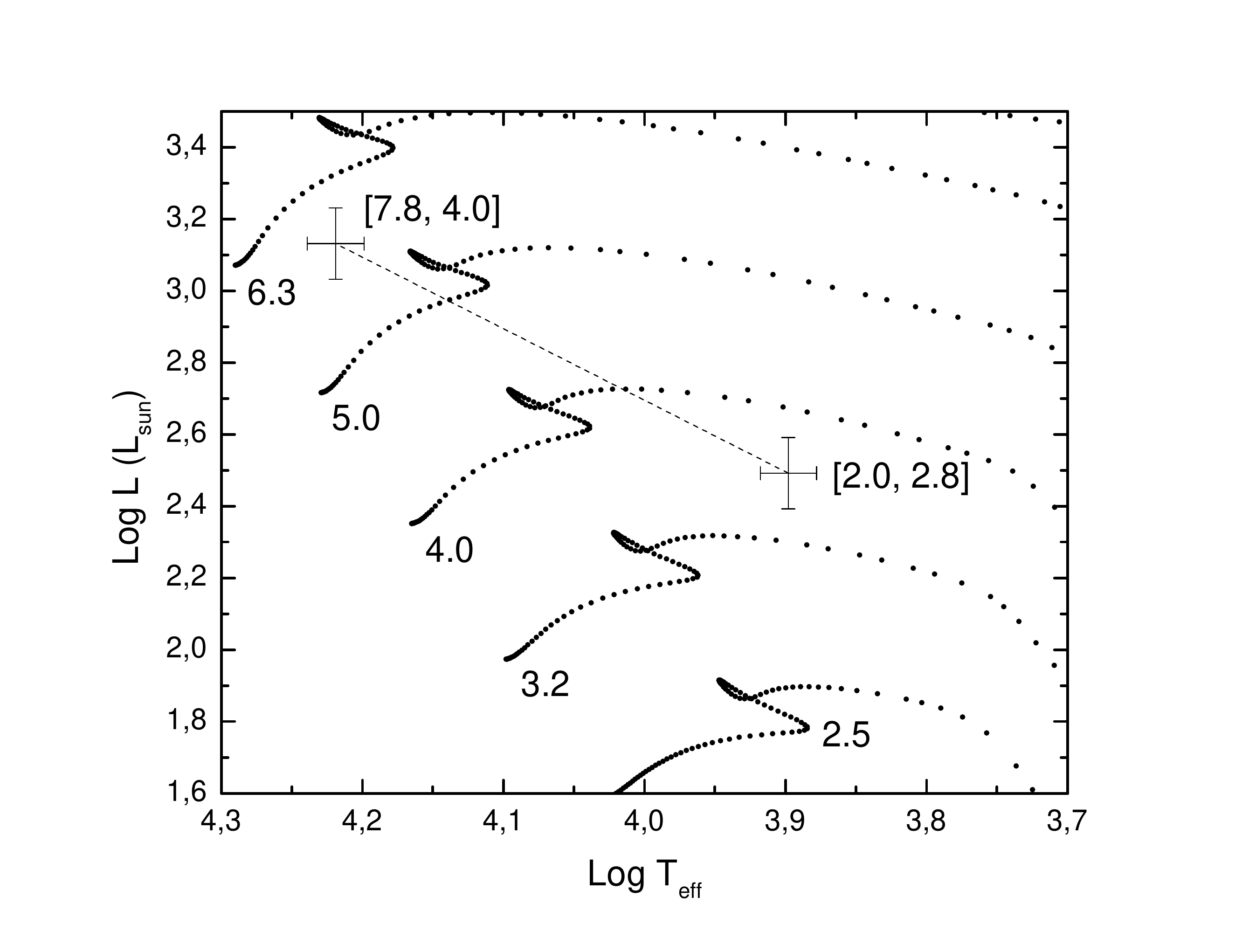}}
\caption{Evolutionary tracks for single stars with solar metallicity (Claret 2004) and the position for the stellar components of V\,393 Sco. Derived masses and log $g$ values are given between parenthesis. The  evolutionary tracks are labeled with initial masses.
}
\label{x}
\end{figure}

\begin{figure}
\scalebox{1}[1]{\includegraphics[angle=0,width=8.5cm]{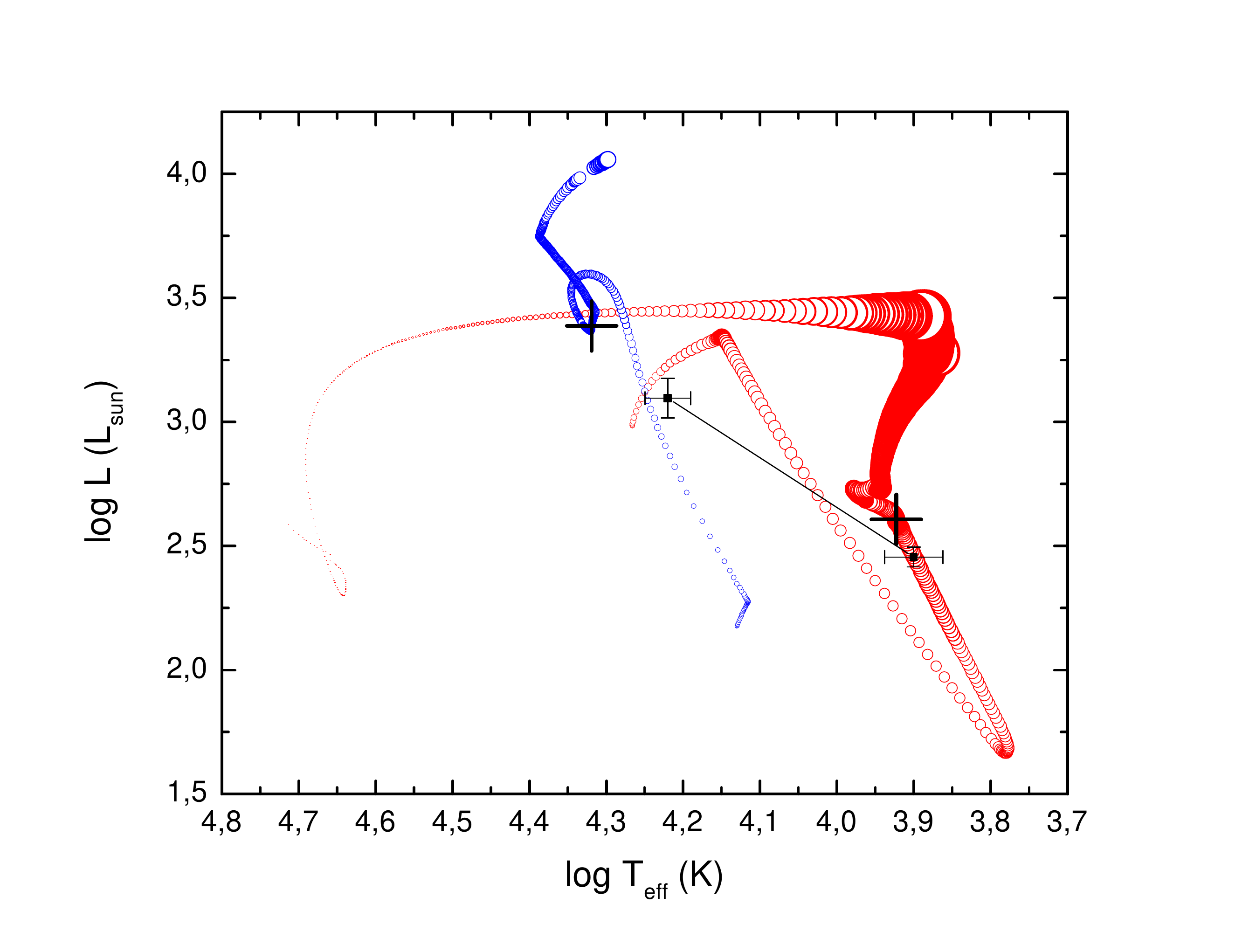}}
\caption{Evolutionary tracks for the  binary star model from Rensbergen et al. (2008) that best fit the data. Donor (right track) and gainer (left track) evolutionary paths are shown, 
along with the observations for V\,393 Scorpii (with error bars and connecting line).
The best fit is reached at the time corresponding to the model indicated by large crosses, that is characterized in Table 3. The mismatch for the primary is discussed in the text. Stellar 
sizes are proportional to the circle diameters.
}
  \label{x}
\end{figure}

\begin{figure}
\scalebox{1}[1]{\includegraphics[angle=0,width=8.5cm]{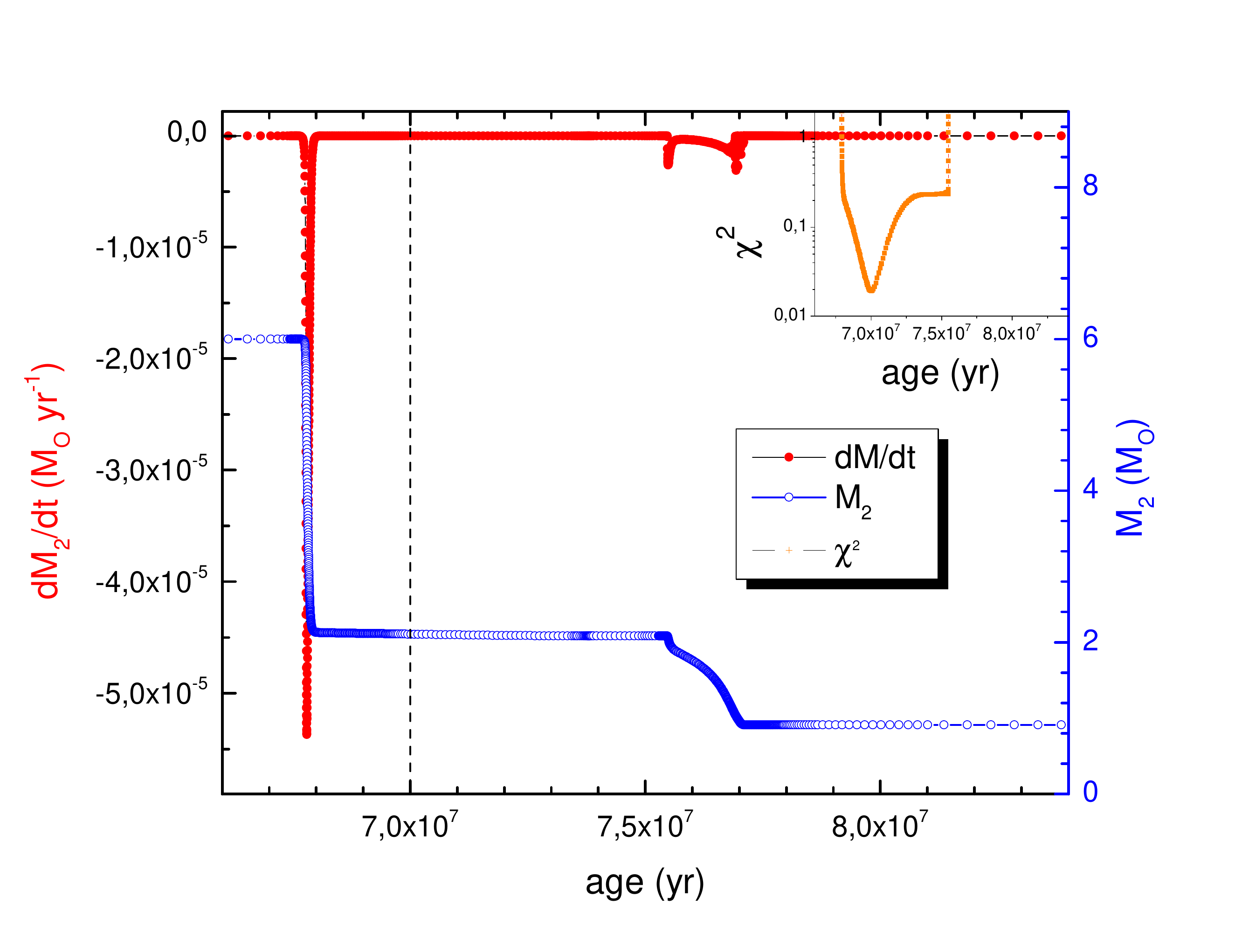}}
\caption{
$\dot{M_2}$ (upper curve) and $M_{2}$ for the best evolutionary model. The vertical dashed line  indicates the position for the best model. $\chi^{2}$ is shown in the inset graph.
}
  \label{x}
\end{figure}

\begin{figure}
\scalebox{1}[1]{\includegraphics[angle=0,width=8.5cm]{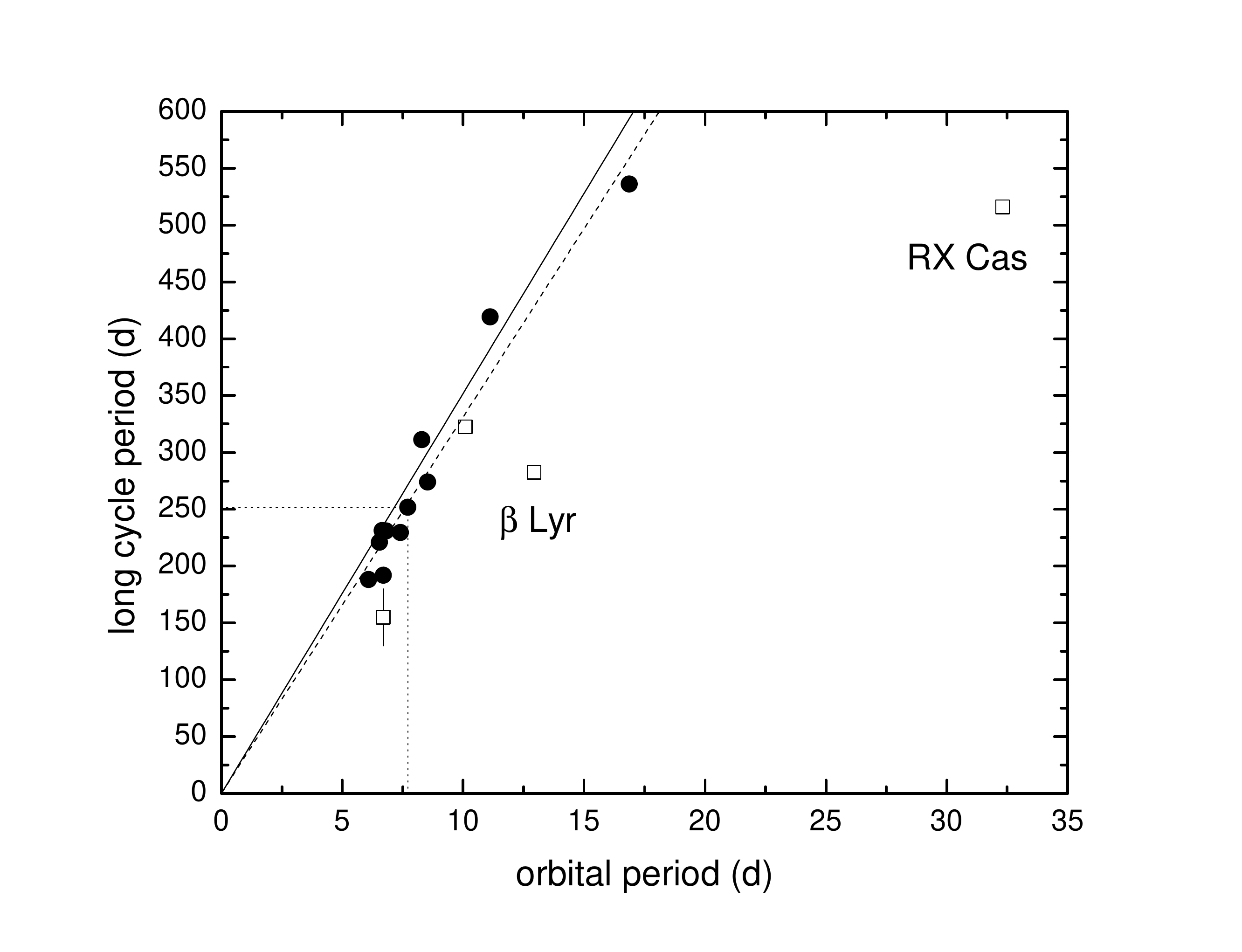}}
\caption{The Double Periodic Variables listed in Table 4 (solid dots) and the hot emission-line binaries with cyclic long-term brightness changes listed in Table 5 (open squares). The upper and lower lines represent the best fit for 30 SMC DPVs (Mennickent et al. 2003) and 125 LMC DPVs (Poleski et al.  2010). The position of V\,393 Scorpii is indicated by dotted lines. }
  \label{x}
\end{figure}

\subsection{Double Periodic Variables in the Milky Way}

We performed a search for Galactic DPVs in the ASAS catalog 
using the software described in Section 3. Light curves were inspected for the presence of two periodicities. We found the 11 systems shown in Table 4 (see also Mennickent et al. 2009 and Michalska et al. 2010 for previous reports on this investigation).  Two of our detected DPVs were previously cataloged as Algol type variables with additional long-term variability, viz.\, AU Mon and V393 Sco. The lack of dedicated stellar variability surveys in our Galaxy probably explains why the long photometric variability of many of these DPVs remained  undetected for a long time. In Table 5 we also  show those Galactic hot emission-line binaries with cyclic long-term brightness changes reported by Desmet et al. (2010, their Table 10).  

The hallmark of the DPV phenomenon, i.e. the correlation between the orbital and the long cycle is shown in Fig.\,6. 
V\,393 Sco fits very well the general tendency. However, two of the objects listed by Desmet et al. (2010), namely 
$\beta$ Lyr and RX\,Cas clearly deviate from it. Excluding these two objects, the best linear fit for the remaining 13 systems is: \\

$P_{long} = 32.7(9) P_{o}$ \hfill(12)\\

\noindent
This relation is comparable to previous relationships found for LMC and SMC DPVs, with respective coefficients 33.1 and 32.4 (Desmet et al. 2011 and Mennickent et al. 2005).

According to this relationship, the predicted long period for $\beta$ Lyr is about 423 days, 1.50 times longer than the reported 282 days, and for
RX\,Cas is 1057 days, i.e. 2.05 times longer than reported. The commensurability of predicted and observed periods is notable. In order to investigate if the expected
423 d periodicity is present in the $\beta$ Lyr  photometry, we removed the orbital variability from the $V$-band light curve kindly provided by Dr. Harmanec and searched for additional frequencies. The residual periodogram shows  the already known periods of 282 and 356 d but only a very small peak in 426 days. Residuals folded with this period show a very noisy diagram.

It is possible that a different evolutionary stage explains the position of the outliers $\beta$ Lyr and RX Cas in Fig.\,6.  These objects could still be inside the burst of mass transfer illustrated in Fig.\,5, while DPVs should already have passed this stage. In favor of this interpretation is the fact that orbital period changes have been detected in $\beta$ Lyr and RX Cas  and interpreted as due to very large mass transfer rates ($\sim$ 10$^{-5}$ M$_{\odot}$ yr$^{-1}$). However, orbital period changes have not been reported in DPVs.

\begin{table*}
 \caption{Double-periodic variables in the Milky Way detected from the analysis of the ASAS database. Remarks are from SIMBAD (http://simbad.u-strasbg.fr/simbad/) except for V\,393 Sco that are from this paper. 
 }
 \begin{tabular}{rrcrrrrrrc}
\hline
 \multicolumn{3}{c}{Names}              &RA(2000)   &  DEC(2000) &  $P_{orb}$  & $P_{long}$&   \multicolumn{2}{c}{TYCHO-2}      & Remarks      \\
HD     &  TYCHO     &GCVS    &           &            &  (d)       &(d)      &   $B_{T}$  &  $V_{T}$     &                \\
\hline
50526  &161-1014-1  & -      &06:54:02.04& +06:48:48.5&  6.7015& 191.7&   8.342&  8.244& B9             \\
50846  &54801-1012-1& AU\,Mon &06:54:54.71& -01:22:32.9& 11.1132& 419.0&   8.453&  8.432&   Ecl B3V+F8IIIe \\
 -     &5978-472-1  & -      &07:26:41.41& -22:08:53.7&  8.2962& 311.0&  10.641& 10.585&        -          \\
 -     &5985-958-1  & -      &07:44:15.30& -17:58:45.6&  7.4062& 229.5&  10.713& 10.641&      -            \\
 -     &8175-333-1  &DQ\,Vel  &09:30:34.22& -50:11:54.0&  6.0833& 188.0&  11.473& 10.977&  Ecl            \\
90834  &   -        & -      &10:27:41.61& -59:17:04.9&  6.8148& 230.7&   9.591&  9.500&    B5III/IVe      \\
135938 &8695-2281-1 & -      &15:20:08.44& -53:45:46.5&  6.6477& 231.2&   9.470&  9.267&    B5/B6IVp       \\
-      &8708-412-1  &GK\,Nor  &15:34:50.92& -58:23:59.1&  6.5397& 221.0&  11.646& 11.275&   Ecl            \\
328568 &8325-3366-1 &LP\,Ara  &16:40:01.78& -46:39:34.9&  8.5331& 274.0&  10.541& 10.243&   Ecl B8         \\
161741 &7385-1101-1 &V393\,Sco &17:48:47.60& -35:03:25.6&  7.7126& 251.7&   7.748&  7.609&  Ecl B4V/A7III     \\
170582 &5703-2382-1 & -      &18:30:47.53& -14:47:27.8& 16.872 & 536.0&  10.150&  9.711&   A3-A9          \\
\hline
\end{tabular}
\end{table*}

\begin{table*}
 \caption{Galactic hot emission-line binaries with cyclic long-term brightness changes (Desmet et al. 2010). $P_{th}$ is the predicted long period according to  Eq.\,12. References for the spectral types are  given.}
 \begin{tabular}{rrcrrrrrc}
\hline
 \multicolumn{3}{c}{Names}              &RA(2000)   &  DEC(2000) &  $P_{orb}$  & $P_{long}$&  $P_{th}$ & Remarks      \\
HD     &  TYCHO     &GCVS       &                  &                     &(d)           & (d)     &(d)    &                  \\
\hline
    -    &4313-258-1 & RX Cas&03:07:45.75& +67:34:38.6 & 32.312&516 &1057 &A5III+G3III; Strupat 1987\\         
174237 &3918-1829-1& CX Dra &18:46:43.09& +52:59:16.7& 6.696 &130-180 &219 &B2.5V+F5III; Simon 1996 \\ 
174638&2642-2929-1&$\beta$ Lyr &18:50:04.80 &+33:21:45.6 & 12.94 &282.4&423 &B8epII+B6.5;  Budding et al. 2004\\ 
216200&3223-3619-1&V360 Lac &22:50:21.77 &+41:57:12.2& 10.085 &322.2  &330 & Be+F; Linnell et al. 2006\\
\hline
\end{tabular}
\end{table*}

\section{Conclusions}

In this paper we have modeled the {\it orbital} light curve of the intermediate-mass interacting binary
\var  to obtain stellar and system parameters. We also disentangled the long-term light curve 
at optical and infrared photometric bands. We have found insights on the system evolutionary stage and  long cycle nature. The main results of our research are:\\

\begin{itemize}
\item The orbital period change, if present, is shorter than 0.5 sec per year.
\item The long-term light curves  are characterized by a smooth oscillation in a time scale of 253 days with larger amplitudes in redder bandpasses. 
\item The best fit to the orbital light curve requires a non-stellar component that was modeled with an optically thick disc model. The disc radius is about half of the Roche lobe radius of the gainer and two bright spots are required to fit the observations. 
\item We found the stellar and system parameters that best match the observations,  which are given in Table 2 along with parameters for the  disc and the bright spots. 
\item The stability of the orbital light curve suggests that the stellar $+$ disc configuration remains stable during the long cycle. Variability of the optically thick disc is not the main source for long cycle.
\item Therefore and in order to fit the redder color at long maximum, we argue that the long cycle is produced by free-free emission in a variable structure, probably visible perpendicular to the orbital plane, something reminiscent of the jets found in $\beta$ Lyr (Harmanec et al. 1996). The suggestion of equatorial mass loss as the cause of the long cycle by M10 is probably  biased by detection of equatorial outflows not necessarily related to the long cycle. 
\item A comparison with published evolutionary tracks provides an estimate for the  age of the system, namely log t =  7 $\times$ 10$^{7}$  yr. We find the system after a mass exchange episode, where 4 M$_{\odot}$ were transferred from the donor to the gainer in a period of 400.000 years. 
\item The evolutionary  model with initial stellar masses of 6 and 3.6 M$_{\odot}$ reproduces relatively well the present donor parameters and orbital period, but overestimates the gainer temperature and luminosity, a fact that could be ascribed to the optically thick disc not considered in the evolutionary tracks. In order to explain these features, we argue that part of the mass, maybe up to 2  M$_{\odot}$ has not been accreted by the gainer, but remains in the {\it massive} optically-thick disc. 
\item We find a discrepancy between  values of $\dot{M}$
 derived from theoretical model fitting of observationally-derived parameters and those derived from accretion-theory analysis of observationally-derived parameters,  the former being smaller by 2 orders of magnitude. If the optically thick disc is a representation of a  massive  disc-like pseudo-photosphere, then its luminosity could not be accretion-driven. 
 The constancy of the orbital period  supports the view that the disk luminosity is not driven by viscosity, but probably by reprocessed stellar radiation.
\item We present the results of our search for Galactic DPVs and make a comparison  with hot emission-line binaries with cyclic long-term brightness changes.  13 Galactic DPVs show a similar correlation between  $P_{o}$  and $P_{long}$   to  that observed in LMC and SMC DPVs. The systems $\beta$ Lyr and RX Cas  deviate from this tendency. These systems could be in an earlier 
evolutionary stage compared with DPVs.

\end{itemize}

\section{Acknowledgments}
 We thank the anonymous referee for useful comments on the first version of this manuscript.  REM acknowledges support by Fondecyt grant 1070705, 1110347, the Chilean 
Center for Astrophysics FONDAP 15010003 and  from the BASAL
Centro de Astrofisica y Tecnologias Afines (CATA) PFB--06/2007. G.D. 
acknowledges the financial support of the Ministry of Education and Science of the Republic of Serbia through the project 176004 ``Stellar physics''. We  thank Nicki Mennekens for conversations about the evolutionary models discussed in this article and Dr. P. Harmanec for kindly  providing the light curve of $\beta$ Lyr for its inspection.


\bsp 
\label{lastpage}
\end{document}